\documentstyle[psfig]{mn}

\newif\ifAMStwofonts

\def\ulum{{\rm erg}\, {\rm s}^{-1}}
\def\uflux{{\rm erg}\, {\rm cm}^{-2}\, {\rm s}^{-1}}
\def\ucol{{\rm cm}^{-2}}
\def\ROSAT{{\it ROSAT}}
\def\ASCA{{\it ASCA}}
\def\kms{{\rm km}\, {\rm s}^{-1}}
\def\keV{{\rm keV}}

\def\dof{{\nu}}

\ifoldfss
  \ifCUPmtlplainloaded \else
    \NewTextAlphabet{textbfit} {cmbxti10} {}
    \NewTextAlphabet{textbfss} {cmssbx10} {}
    \NewMathAlphabet{mathbfit} {cmbxti10} {} 
    \NewMathAlphabet{mathbfss} {cmssbx10} {} 
  \fi
  \ifAMStwofonts
    \ifCUPmtlplainloaded \else
      \NewSymbolFont{upmath} {eurm10}
      \NewSymbolFont{AMSa} {msam10}
      \NewMathSymbol{\upi}     {0}{upmath}{19}
      \NewMathSymbol{\umu}     {0}{upmath}{16}
      \NewMathSymbol{\upartial}{0}{upmath}{40}
      \NewMathSymbol{\leqslant}{3}{AMSa}{36}
      \NewMathSymbol{\geqslant}{3}{AMSa}{3E}

    \fi
  \fi
\fi 

\ifnfssone
  \newmathalphabet{\mathit}
  \addtoversion{normal}{\mathit}{cmr}{m}{it}
  \addtoversion{bold}{\mathit}{cmr}{bx}{it}
  \newmathalphabet{\mathbfit} 
  \addtoversion{normal}{\mathbfit}{cmr}{bx}{it}
  \addtoversion{bold}{\mathbfit}{cmr}{bx}{it}
  \newmathalphabet{\mathbfss} 
  \addtoversion{normal}{\mathbfss}{cmss}{bx}{n}
  \addtoversion{bold}{\mathbfss}{cmss}{bx}{n}
  \ifAMStwofonts
    \ifCUPmtlplainloaded \else
      \UseAMStwoboldmath
      \makeatletter
      \new@mathgroup\upmath@group
      \define@mathgroup\mv@normal\upmath@group{eur}{m}{n}
      \define@mathgroup\mv@bold\upmath@group{eur}{b}{n}
      \edef\UPM{\hexnumber\upmath@group}
      \new@mathgroup\amsa@group
      \define@mathgroup\mv@normal\amsa@group{msa}{m}{n}
      \define@mathgroup\mv@bold\amsa@group{msa}{m}{n}
      \edef\AMSa{\hexnumber\amsa@group}
      \makeatother
      \mathchardef\upi="0\UPM19
      \mathchardef\umu="0\UPM16
      \mathchardef\upartial="0\UPM40
      \mathchardef\leqslant="3\AMSa36
      \mathchardef\geqslant="3\AMSa3E
    \fi
  \fi
\fi 

\ifnfsstwo
  \DeclareMathAlphabet{\mathbfit}{OT1}{cmr}{bx}{it}
  \SetMathAlphabet\mathbfit{bold}{OT1}{cmr}{bx}{it}
  \DeclareMathAlphabet{\mathbfss}{OT1}{cmss}{bx}{n}
  \SetMathAlphabet\mathbfss{bold}{OT1}{cmss}{bx}{n}
  \ifAMStwofonts
    \ifCUPmtlplainloaded \else
      \DeclareSymbolFont{UPM}{U}{eur}{m}{n}
      \SetSymbolFont{UPM}{bold}{U}{eur}{b}{n}
      \DeclareSymbolFont{AMSa}{U}{msa}{m}{n}
      \DeclareMathSymbol{\upi}{0}{UPM}{"19}
      \DeclareMathSymbol{\umu}{0}{UPM}{"16}
      \DeclareMathSymbol{\upartial}{0}{UPM}{"40}
      \DeclareMathSymbol{\leqslant}{3}{AMSa}{"36}
      \DeclareMathSymbol{\geqslant}{3}{AMSa}{"3E}
    \fi
  \fi
\fi 

\ifCUPmtlplainloaded \else
  \ifAMStwofonts \else 
    \def\upi{\pi}
    \def\umu{\mu}
    \def\upartial{\partial}
  \fi
\fi

\title[A radio-loud obscured AGN at z=1.246]{Discovery of an
X-ray selected radio-loud obscured AGN at
z=1.246}
\author[X. Barcons et al.]
       {X. Barcons$^1$, R. Carballo$^{1,2}$, M.T. Ceballos$^1$, 
R.S. Warwick$^3$, J.I. Gonz\'alez-Serrano$^1$ \\
        $^1$ Instituto de F\'\i sica de Cantabria (Consejo Superior de
Investigaciones Cient\'\i ficas - Universidad de Cantabria), 39005
Santander, Spain\\
        $^2$ Departamento de F\'\i sica Moderna, Universidad de
Cantabria, 39005 Santander, Spain\\
        $^3$ Department of Physics and
Astronomy, University of Leicester, Leicester LE1 7RH, UK\\}

\date{17 July 1998}

\pagerange{\pageref{firstpage}--\pageref{lastpage}}
\pubyear{1998}

\begin{document}

\maketitle

\label{firstpage}

\begin{abstract}
We have discovered an obscured active galaxy at 
redshift $z = 1.246$ identified with the {\it ROSAT} X-ray source 
RX J1011.2+5545. We report on multiwavelength observations of this
source and discuss its X-ray, optical and radio properties.  This is the 
first X-ray selected, obscured active galaxy at high redshift to be shown to 
be radio-loud, with a
radio counterpart exhibiting a classical double-lobe morphology. 

\end{abstract}

\begin{keywords}
Galaxies: active, X-rays: galaxies, Radio Continuum: galaxies
\end{keywords}

\section{Introduction}

Deep X-ray surveys have recently revealed a population of moderately to heavily
absorbed active galactic nuclei (AGN) at faint fluxes. A few such objects 
are known to be at high redshift, for example one source discovered by
\ROSAT\ is at z=2.35 (Almaini et al 1995) and two others discovered by 
\ASCA\ have z=0.9 and z=0.672 (Ohta et al 1996; Boyle et al 1998).  
The so-called Narrow-Line X-ray Emitting Galaxies (NLXGs) might, in fact,
be the low redshift counterparts of these obscured objects, since both
classes are characterised by hard X-ray spectra (Carballo et al 1995; 
Almaini et al 1996).  The discovery of 
such sources at faint X-ray fluxes is of vital importance in explaining the 
origin of the X-ray background, since the brightest AGN in the X-ray sky 
(mostly type 1 AGN) generally have much softer X-ray spectra than the X-ray 
background spectrum (Fabian \& Barcons 1992).

The UK {\it ROSAT} Medium Sensitivity Survey (Branduardi - Raymond et al
1994; Carballo et al 1995) was carried out in order to identify a
complete sample of moderately faint X-ray selected sources (flux over
the 0.5--2 keV band in excess of $1.7\times 10^{-14}\uflux$) over a
significant area of the sky (2.2 deg$^2$) in a region of minimal
Galactic absorption. In this survey the source with the highest hardness
ratio is RX J1011.2+5545 with $HR=0.67$ ($HR=(H-S)/(H+S)$ where $S$ and 
$H$ are the counts in PSPC channels 11--39 and 40--200 respectively). It is
also one of its brightest sources with a flux $S(0.5-2\,
\keV )=6.6\times 10^{-14}\, \uflux$. The hard X-ray spectrum together
with the fact that the source has no optical counterpart visible on the POSS
plates (which is atypical of the X-ray sources at this flux level), 
suggested a possibly highly obscured source and prompted us to start a
program of follow-up optical and  \ASCA\ hard X-ray observations. A
NED search also revealed that the source is a radio-emitter at various
frequencies, with a double lobe morphology.  The combination of
radio, optical and X-ray data has enabled us to classify this object as a
radio-loud, moderately obscured, high-excitation AGN at a redshift
$z=1.246$. This is the first X-ray selected obscured AGN discovered 
at high redshift found to be radio loud. In this paper we report on all 
of the recent observations and discuss the nature of this source.

\section {The Data}

\subsection{{\it ROSAT} soft X-ray observations}

The discovery observation was carried out on May 11, 1992 with 
the \ROSAT\ PSPC-B, giving an exposure time of 18529s. The data were 
reduced and scanned for sources as described in Carballo et al (1995). 
After a number of sources in the PSPC image were identified with optical 
counterparts, the astrometry of the X-ray field was corrected by applying 
shifts in RA and DEC. The final X-ray position for the source  
RX J1011.2+5545 is  $10^h11^m12^s.4$ and $55^{\circ}44'50''$
(J2000) with a  90 per cent error circle of radius $\sim 4''$.
The X-ray image showed no evidence for any extension in the source
(the FWHM is $27''$, consistent with the PSF at an offset angle
from the \ROSAT\ field centre of  $7.3'$).
The Galactic column density in this direction is $6.7\times 10^{19}\, {\rm
cm}^{-2}$.

%
%
%
%
%
%

 We used the FTOOLS/XSELECT V3.6 package to extract the counts contained
 within a circle of radius $1.5'$ centered on the source and used 
 a ``source-free'' region of radius  $6.5'$ at a similar off-axis angle
 in the background subtraction. For the purpose of spectral fitting we 
 grouped the PSPC pulse-height data so that every spectral bin
 contained at least 20 counts, leading to a 0.1--2.0 keV source
 spectrum with just 6 bins. 

 A single power-law fit, assuming only Galactic line-of-sight absorption, 
 gives a very flat photon spectral index
 $\Gamma=0.93_{-0.23}^{+0.20}$ (we always quote 90 per cent errors for
 a single parameter). However, the quality of the fit is not very good
 ($\chi^2/\dof=10.2/4$) corresponding to probability for the null 
 hypothesis (PNH) of only 3.7 per cent.  The inclusion in the spectral model
 of absorption 
 intrinsic to the X-ray source produces a somewhat better fit 
 ($\chi^2/\dof=4.4/3$) with a steeper underlying power law
 (although formally the improvement in the fit is not significant
 in terms of the F-test). Clearly data at higher energies are required
 in order to better constrain the continuum slope in this source.

\subsection{{\it ASCA} hard X-ray observations}

RX J1011.2+5545  was observed with \ASCA\ on November 12-13,
1995. The source was clearly seen in both the SIS0 and SIS1 cameras,
which were operated in 1-CCD mode (Bright mode).  Standard FTOOLS/XSELECT V3.6
tasks and techniques were used to clean the data (using default
parameters), resulting in effective exposure times of 53054s (SIS0)
and 52855s (SIS1).  In this paper we ignore the GIS2 and GIS3
observations since the source is barely detected in these detectors. 
We rely on the spectral calibration of the SIS0 data, in preference to that 
for SIS1 when necessary.

A spectrum was extracted from a $3'$ radius region centred on the
source.  The background subtraction was found to be much more accurate
when we chose a source-free region within the same image rather than
using the available archival background images. (For example, a
detector Fe fluorescent line in the 6-7 keV spectral region went away
when we used the adopted method, but not when the archival background
was used).  After background subtraction the resulting spectrum was
again binned in order to give a minimum of 20 counts per spectral
channel.  The result was 16 bins for SIS0 and 15 for SIS1. No
significant source variability was found in the data.

\begin{figure}
\centerline{\psfig{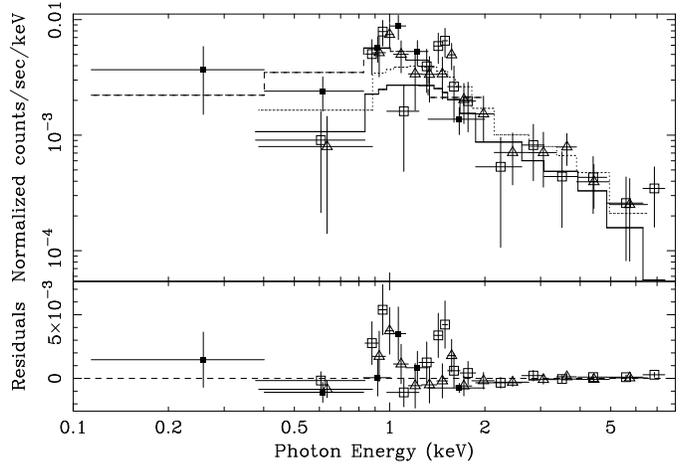}}
 \caption{The measured \ROSAT\ and \ASCA\ X-ray spectra of RX J1011.2+5545
 together with the residuals to the best fitting (power-law plus
 intrinsic absorption) model. The filled squares, empty squares and triangles
 are the \ROSAT\ PSPC, the \ASCA\ SIS0 and the \ASCA\ SIS1 data points
 respectively. The \ROSAT\ PSPC model is shown with a dashed line, the
 \ASCA\ SIS0 one with a solid line and the \ASCA\ SIS1 one with a
 dotted line.}
\end{figure}

The simultaneous fitting of the SIS0 and SIS1 data with a single power-law 
model (but with different normalisations applying to the two detectors
to allow for calibration uncertainties) gives an acceptable fit
($\chi^2/\dof=34.5/28$ with PNH of 18.5 per cent) with a rather flat photon
index $\Gamma=1.43_{-0.23}^{+0.24}$. The inclusion of absorption
intrinsic to the X-ray source again produces a steeper underlying power law
but with only a modest improvement in the fit ($\chi^2/\dof=31.8/27$)
(which again is not a significant improvement in terms of the F-test).

\begin{figure}
\centerline{\psfig{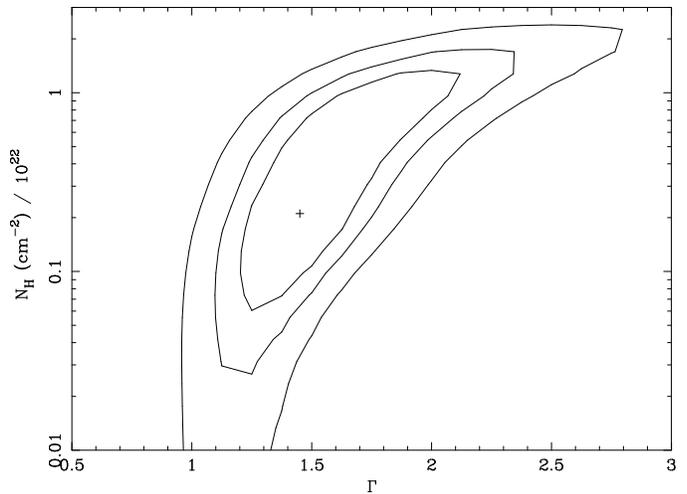}}
 \caption{Confidence contours (68, 90 and 99 per cent confidence) for the intrinsic absorption and photon
 index from the combined \ROSAT\ and \ASCA\ data. }
\end{figure}

We then combined the \ROSAT\ and \ASCA\ data so as to better constrain the
spectral parameters.  Our approach has been  to assume the same value of the
model normalisation for the \ROSAT\ and \ASCA\ SIS0 data but allow
a different normalisation for \ASCA\ SIS1 data.  This procedure
produces a significantly better fit than taking the same normalisation for all
three datasets (at 99.9 per cent using F-test), whereas introducing different
normalisations for each instrument does not result in a
significant improvement.  A single power law fit is only marginally
acceptable ($\chi^2/\dof=53.9/34$ with a PNH=1.6 per cent), with
$\Gamma=1.13\pm 0.16$.  However, the fit improves if absorption intrinsic to 
the source is included ($\chi^2/\dof=47.2/33$ with PNH of 5.2 per cent).  
The best fit (see Fig. 1) corresponds to $\Gamma=1.45^{+0.72}_{-0.28}$ and
$N_H=(2.1^{+12.4}_{-1.6})\times 10^{21}\, \ucol$ (at the redshift of
the source $z=1.246$). Fig. 2 shows the confidence
contours for the two free spectral parameters.

There is some evidence for significant residuals in all three
instruments at $\sim 1$ $\keV$ (Fig. 1). The inclusion of a
Gaussian-line component centered at this energy results in a further
improvement of the fit ($\chi^2/\dof= 31.8/29$). However, it is not
completely obvious that such a feature is associated with the source;
the corresponding rest-frame energy is $2.2\pm 0.1\, \keV$ with a
rest-frame equivalent width $\sim 165\, {\rm eV}$. One could identify
this as a SiXIV-SiXVI complex (Netzer \& Turner 1997), but then the Fe
K line should be seen in the spectrum and it is not (rest-frame
equivalent width $< 418$ eV at 95 per cent confidence).  Attempts to
account for these residuals in the X-ray spectrum in terms of ionised
absorbers did not show significant improvemet in the fit.

We conclude that the absorbed, power-law fit is the most tenable model.
The flux of the source is $S(0.5-2\, \keV)=6.6\times 10^{-14}\, \uflux$ and
$S(2-10\, \keV)=1.9\times 10^{-13}\, \uflux$ and the K-corrected rest
frame luminosity (using the measured redshift of $z=1.246$) is
$L(0.5-2\, \keV)= 4.8\times 10^{44} \ulum$ and $L(2-10\, \keV)=
2.1\times 10^{45} \ulum$ ($H_0=50\, {\rm km}\, {\rm s}^{-1}\, {\rm
Mpc}^{-1}$ and $q_0=0$).

\subsection{Optical imaging}

The POSS plates show no counterpart within or near the position of the
\ROSAT\ source. In order to search for fainter candidate optical 
counterparts, we imaged the field (as for all the other survey sources) 
with the CCD
camera at the Cassegrain focus of the 2.2m telescope of the Centro
Astron\'omico Hispano Alem\'an, Calar Alto on February 10,
1994. A single exposure of 900s was taken with the Johnson R filter.
The photometric conditions were good and the seeing was
$\sim 1.4''$. Data reduction and the astrometric and photometric 
calibrations were performed as described by Carballo 
et al (1995).

\begin{figure}
\centerline{\psfig{figure=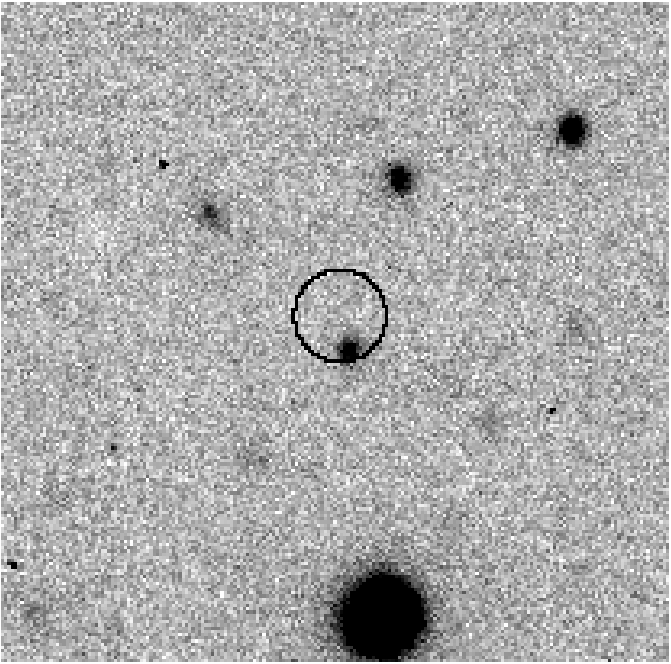,width=0.5\textwidth,angle=0}}
 \caption{$R$-band image of the field around RX J1011.2+5545. The
 image is 1 arcmin each side. The
 circle is the 90 per cent error circle for the \ROSAT\ X-ray position}
\end{figure}

The R-band image (Fig. 3) reveals a single $R=21.02\pm 0.08$ source
within or near the error circle of the X-ray source, whose position is
$10^h11^m12^s.3$ and $55^{\circ}44'47''$ (J2000) and is the likely
counterpart, later confirmed by spectroscopy. The surface brightness
profile of the source does not show compelling evidence for any
additional extension to the profile of a bright star.  

\subsection{Optical spectroscopy}

Optical spectroscopy of the candidate counterpart was carried out at
the 4.2m William Herschel Telescope on the Observatorio del Roque de
los Muchachos (La Palma) with the ISIS double spectrograph, on 
February 25, 1998. We used the 150 lines/mm gratings and TEK CCD
detectors for both arms, covering a spectral range from 3400 to 8550\AA .
The atmospheric conditions were very poor with bad and variable sky
transparency, dust and seeing, with the latter starting at $3.5''$ but 
later improving to between $2''$ and $2.5''$. Two sets of observations were 
carried out, the first set corresponding to the period of worst seeing 
with a slit width of $2.5''$  and the second set with a slit width of 
$1.5''$.  Here we ignore the first set of observations, although 
qualitatively they reveal much the same as the second set. 

\begin{figure}
\centerline{\psfig{figure=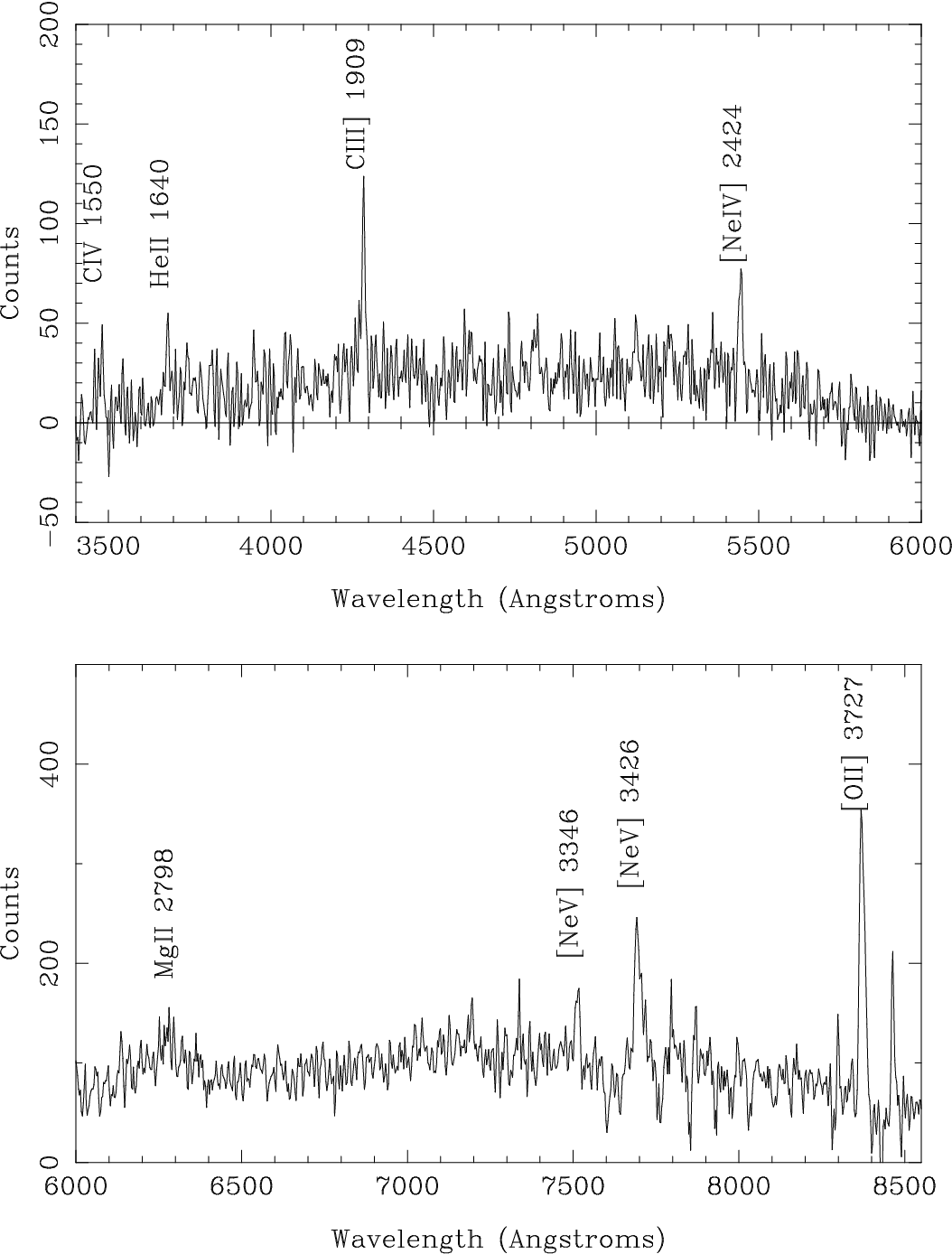,width=0.5\textwidth,angle=0}}
 \caption{Raw optical spectrum of RX J1011.2+5545.}
\end{figure}

The observations with the  slit width set at $1.5''$ totalled 5 on-source
exposures of 1800s each, all close to parallactic angle and
with airmass less than 1.2. The data were reduced using standard IRAF
routines. The optimally extracted source spectra were registered to a common
wavelength origin using the sky spectrum.  The resulting summed
spectrum was wavelength calibrated using polynomial fits to standard
arc maps, yielding rms residuals of 0.72\AA\ and 0.37\AA\ in the blue
and in the red respectively. The spectral resolution was measured from
unblended arc lines to be 9.6\AA\ and 8.8\AA\ in the blue and in the
red respectively.  Given the poor conditions, no attempt
was made to flux calibrate the spectra.

Fig. 4 shows the resulting spectra with markers on the most prominent
emission lines.  The redshift $z=1.246$ has been determined from the
strongest features [NeV]$\lambda$3426 and [OII]$\lambda$3727, although
the other emission lines are entirely consistent with this
redshift. The presence of the high ionisation [NeV]$\lambda$3346 and
$\lambda$3426 lines clearly reveals an AGN. Table 1 lists the emission
features detected in the spectrum, rest-frame equivalent widths and
FWHM estimated via gaussian fitting with 90 per cent errors and
corrected for spectral dispersion.

\begin{table}
\centering
\begin{minipage}{70mm}
\caption{Detected emission lines in the optical spectrum}
\begin{tabular}{llcc}

\hline

Emission & Redshift & $W_{\lambda}^a$ & FWHM\\
line &           & (\AA ) & ($\kms$)\\

\hline
CIV$\lambda$1550       & 1.2450    & 35$^b$ & $<2500^b$\\
HeII$\lambda$1640      & 1.2452    & 15$^b$ & $<800^b$\\
CIII$]\lambda$1909   & 1.2445    & 16     & $385^{+390}_{-380}$\\
$[$NeIV$]\lambda$2423& 1.2469    & 12     & $560^{+280}_{-270}$ \\
MgII$\lambda$2798      & 1.242$^b$ & 15$^b$ & $2000^{+2700}_{-1000}$ ($^b$)\\
$[$NeV$]\lambda$3346 & 1.2453    & 4      & $480^{+160}_{-130}$\\
$[$NeV$]\lambda$3426 & 1.2462    & 14     & $920^{+310}_{-240}$\\
$[$OII$]\lambda$3727 & 1.2462    & 42     & $625^{+60}_{-55}$\\ \hline
\end{tabular}

$^a$ Rest-frame equivalent width\\
$^b$ Highly uncertain\\
\end{minipage}
\end{table}

The semi-forbidden CIII] line is clearly detected and narrow (see
Table 1).  Since this line is predicted to be broad in a type 1 AGN,
the implication is that the broad-line region in this AGN is
obscured. The CIV and HeII lines appear also narrow, but in a low
signal to noise part of the spectrum. The MgII line is probably broad,
but with an equivalent width normalised to the equivalent width of the
narrow lines significantly smaller (10--20 times) than is typically
found in type 1 AGN (Francis et al 1991).  Broad MgII has been found
in IR hyperluminous galaxies (Hines \& Wills 1993; Hines et al 1995)
and high-redshift radiogalaxies (di Serego Alighieri, Cimati \&
Fosbury 1994; Stockton, Kellogg \& Ridgway 1995) and has been
interpreted as scattered emission from a hidden type 1 AGN.

\subsection{Radio data}

We searched in various archives for radio observations of our source.
There are a number of detections, the most relevant of which are the
Westerbork Northern Sky Survey (Rengelink et al 1997) at 326 MHz, the
Texas Survey (Douglas et al 1996) at 365 MHz, the FIRST survey (White
et al 1997) at 1.4 GHz and the Green Bank 6cm survey (Gregory et al
1996) at 4.85 GHz.  Both the Texas and the FIRST surveys resolve the
source into two components aligned approximately N-S. The N component is
the brightest in the FIRST data (0.090 Jy compared to the 0.071 Jy of
the S component). The optical position lies in between both
components (see Fig. 5). The separation between the components 
is $\sim 11''$ at 1.4 GHz and $\sim 15''$ at 365 MHz.

\begin{figure}
\centerline{\psfig{figure=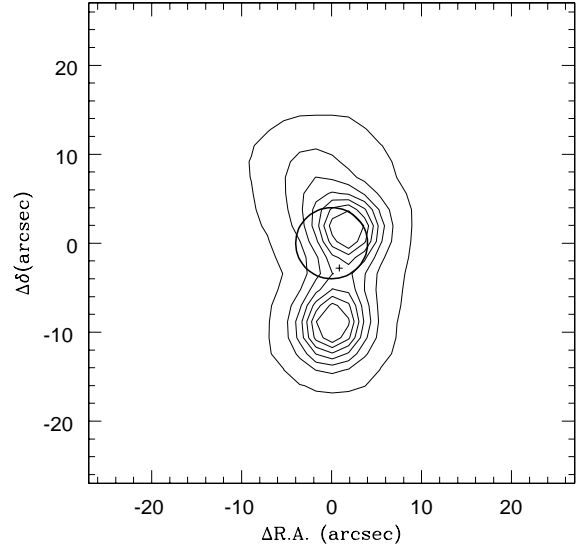,width=0.5\textwidth,angle=0}}
 \caption{A radio map of RX J1011.2+5545 at 1.4GHz from the FIRST
 survey. The cross shows the position of the optical source and the
 thick circle is the error circle of the X-ray source.}
\end{figure}

The integrated radio fluxes  together with the measurements at optical and 
X-ray frequencies are shown in Fig. 6 in the form of a spectral
energy distribution.  The radio spectrum has a 
$S_{\nu}\propto \nu^{-0.9}$ shape from 326 MHz to 4.85 GHz, which is
typical of lobe-dominated radio sources. Although
from the spatial information at the various frequencies it is not
completely clear that this is a lobe-dominated double source, both the
spectral index and the position of the optical source strongly support
this hypothesis.

\begin{figure}
\centerline{\psfig{figure=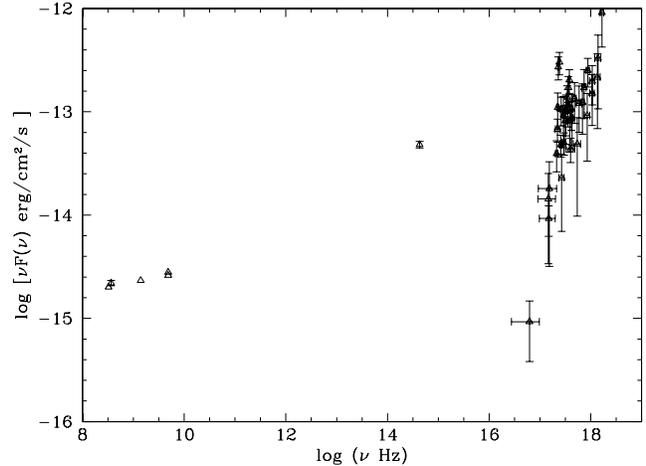,width=0.5\textwidth,angle=270}}
 \caption{The spectral energy distribution of RX J1011.2+5545 from
 radio to X-rays (see text for details on the data points).}
\end{figure}

\section{Discussion}

There are various facts that support the conclusion that RX
J1011.2+5545 is an AGN, the most relevant being a high radio to optical
flux ratio, a $2-10\, \keV$ luminosity exceeding $10^{45}\, \ulum$ and
the broad MgII emission. In addition the strong [NeV]$\lambda$3624 line 
implies the presence of an underlying hard ionising continuum.

We initially suspected obscuration in this source because of its high 
\ROSAT\ PSPC hardness ratio. Also for a typical uncovered AGN, the average
optical magnitude corresponding to its X-ray flux would be $R\sim 19$ (see,
e.g., Hasinger 1996) instead of the observed value of $R\sim 21$.
The absence of a broad CIII] line confirms this obscuration hypothesis.

A weak broad MgII line is detected, its equivalent width being 3 to 5
times smaller than for a type I AGN (Francis et al 1991; Baker \&
Hunstead 1995).  This cannot be explained as a simple obscuration
effect, since in that case both the broad lines and the nuclear
continuum would be equally suppressed, leaving the equivalent widths
unchanged.  The weakness of MgII and the absence of broad CIV and HeII
may be the result of dilution by a source of blue continuum over and
above that emanating directly from the nucleus. The requirement would
be that at a rest wavelength of $\sim 2800$\AA\ the nuclear continuum
may be only 20 to 50 per cent of the total. The nature of this extra
blue component is unknown, but reflected nuclear radiation, nebular
continuum and copious star formation are all possibilities. The
non-detection of a reflected Fe K line in X-rays and the strong [OII] line
with respect to typical type I situation favour the enhanced star
formation scenario.  The equivalent width of the broad CIII] component
is expected to be roughly 2 to 5 times smaller than that of MgII in a
type I AGN, and therefore it would be very weak in this object.
Obscuration of the nuclear continuum could also lead to the narrow
[NeV] lines having enhanced equivalent widths.

%
%

The power-law in the X-ray spectrum of this object is similar to that
found for other luminous radio-loud quasars at high redshifts,
($\Gamma\sim 1.5$, Cappi et al 1997), distinctively flatter than for
radio-quiet AGN. This has been associated with different emission
mechanisms (synchrotron self-Compton with the radio-emitting electrons
in radio-loud AGN versus nuclear emission in radio-quiet objects). It
is then possible that in radio-loud active galaxies the line-of-sight
to X-ray emitting regions intercepts less obscuring material than does
the direct path to the nucleus. Larger absorbing columns ($N_H\sim
10^{22}\, \ucol$) than that observed in RX J1011.2+5545 are common
only among radio-loud quasars at very high redshifts ($z>3$, Cappi et
al 1997, Fiore et al 1998). The possible contribution to the X-ray
flux from a cluster of galaxies hosting this source (which might be
dominant in radiogalaxies, Crawford \& Fabian 1996) is small, since
the X-ray data does not show evidence for a spectral cutoff consistent
with thermal emission.

The amount of X-ray absorption predicts an optical extinction for the
X-ray source which is $A_V=1.1^{+6.7}_{-0.85}$, using standard dust to
gas ratios. For moderate extinction ($A_V\sim 1-2$), the nuclear light
seen in the optical can be direct radiation from the nucleus.  However, if 
the obscuration is much larger, then the MgII
broad line would be seen through reflection only.  It is even possible
that the nucleus is very heavily obscured in the optical ($A_V\gg 10$) in
which case the direct X-ray continuum and nuclear Fe K emission might also be
suppressed, leaving a dominant X-ray component arising in the radio lobes 
with only moderate associated photoelectric absorption. 
Disentangling both possibilities requires high spatial resolution
optical and IR observations.

In any event, the discovery of this object demonstrates that
high-redshift radio-loud obscured AGN are present at faint X-ray
fluxes. Such objects may play a role, albeit probably minor, in 
producing the X-ray background. Surveys to be carried out with AXAF and XMM 
will undoubtely find large numbers of obscured AGNs and show what is their 
contribution to the X-ray background.

\section*{Acknowledgments}

XB and RC were visiting astronomers of the Centro-Astron\'omico
Hispano-Alem\'an, Calar Alto, operated by the Max-Planck-Institute for
Astronomy, Heidelberg jointly with the Spanish `Comisi\'on Nacional
de Astronom\'\i a'.  The William Herschel Telescope is operated on the
island of La Palma by the Isaac Newton Group in the spanish
Observatorio del Roque de los Muchachos of the Instituto de Astrof\'\i
sica de Canarias.  This research has made use of the NASA/IPAC
Extragalactic Database (NED), which is operated by the Jet Propulsion
Laboratory, California Institute of Technology under contract with the
National Aeronautics and Space Administration. XB, RC, MTC and JIGS
acknowledge financial support by the DGES under project PB95-0122.

\label{lastpage}

\end{document}